\documentclass{cta-author}
\usepackage{subfigure}
\usepackage{bm}

\newcommand\CONF[1]{\bm{x}_{#1}}
\newcommand\MAN[1]{s_{#1}}
\newcommand\NEXT[1]{f_{#1}}

\newcommand\NEI{\mathcal{N}}

\newcommand\ACC{\bm{a}}
\newcommand\SCOST{S}

\newcommand\tbrac[1]{(#1)}

\newcommand{\iest}[3]{\widehat{#1_{i,#2}}\langle#3\rangle}
\newcommand{\ijpred}[3]{\widetilde{#1_{i,j}}\langle#2,#3\rangle}
\newcommand{\ipred}[3]{\widetilde{#1_{i,i}}\langle#2,#3\rangle}
\newcommand{\ijcomp}[2]{\overline{#1_{i,j}}\langle #2 \rangle}
\newcommand{\icomp}[3]{\overline{#1_{i,#2}}\langle #3 \rangle}
\newcommand{\kcomp}[3]{\overline{#1_{#2}}\langle #3 \rangle}
\newcommand{\aggcost}{w^a}

\usepackage{hyperref}
\usepackage{todonotes}
{}
{}
{}

\begin{document}

\title{A Game-Theoretic Approach to Decision Making for Multiple Vehicles at Roundabout}

\author{\au{Sasinee Pruekprasert$^{1*}$}
\au{J\'er\'emy Dubut$^{1,2}$}
\au{Xiaoyi Zhang$^1$}
\au{Chao Huang$^3$}
\au{Masako Kishida$^1$}
}

\address{\add{1}{National Institute of Informatics, Hitotsubashi 2-1-2, Tokyo 101-8430, Japan}
\add{2}{Japanese-French Laboratory for Informatics}
\add{3}{College of information science and engineering, Northeastern University,  110167, China}
\email{sasinee@nii.ac.jp}}

\begin{abstract}
In this paper, we study the decision making of multiple
autonomous vehicles at a roundabout.
The behaviours of the vehicles depend on their \emph{aggressiveness}, which indicates how much they value speed over safety.
We propose a distributed decision-making process
that balances safety and speed of the vehicles.
In the proposed process,
each vehicle
estimates other vehicles' aggressiveness and
formulates the interactions among the vehicles 
as a finite sequential game.
Based on the Nash equilibrium of this game, the vehicle
predicts other vehicles' behaviours and
makes decisions. 
We perform numerical simulations to illustrate the
effectiveness of the proposed process, both for safety (absence of
collisions), and speed (time spent within the roundabout).

\end{abstract}

\maketitle

\section{Introduction}

The demand for safety, energy saving, environmental protection, and
comfortable transportation services has been increasing. Thus, it is a
global consensus to accelerate the
development of autonomous vehicles, which incorporate many advanced
technologies such as smart sensors and
wireless vehicle-to-vehicle communication. For this reason,
governments around the world have begun to develop strategies to address
the challenges that arise from
autonomous driving \cite{kaur2018trust}.  An autonomous vehicle is 
assumed to be capable of sensing its environment, making a decision on driving manoeuvre, planning trajectory candidates, selecting an optimal
trajectory based on a certain cost function and finally, executing
desirable control actions \cite{schwarting2018planning}.

Decision making is the key to autonomous driving as it provides guidance 
on if, how, and when the vehicle should change its driving manoeuvre, for example,
from staying the current lane to changing to the adjacent lane
\cite{schwarting2018planning, Liu2018}. The primary concern for making a
decision is safety \cite{Chentong2018}, such as knowing whether there is a potential 
collision after changing the lane, or whether it is legal to turn right at an intersection. The
decision-making process depends not only on  the vehicle's status but also on the
interactions among various road users \cite{galceran2015multipolicy}.

Rule-based decision-makers were the first approaches applied to autonomous driving. 
The principle of these approaches is to make a decision by utilising an expert system. 
For example, Perez et al. \cite{perez2011longitudinal} designed an overtaking fuzzy decision system on a two-way road using fuzzy logic. In \cite{gipps1986model}, Gipps proposed a decision making for changing lanes  based on multiple critical factors such as safe gap distances and the driver's
intended turning movement. Although the rule-based approaches are suitable for the implementation in a simple and specific scenario, they become unreliable in a complex driving environment involving several traffic rules \cite{wang2019lane,zimmerman2004implementing,gipps1986model}.

Learning-based approaches have been popular since they are more capable than the rule-based
approaches in term of dealing with continuously changing environments  \cite{qiao2018automatically}.
In \cite{liu2019novel}, Liu et al. designed a
decision model for autonomous lane changing based on Gaussian support vector machine.
Reinforcement learning, which is
a well-known class of learning paradigm, shows its promising
benefits in determining optimal decisions in various tasks
\cite{qiao2018automatically,li2015reinforcement, you2018highway,xu2018reinforcement}. 
The combination of reinforcement learning and other methods, such as deep learning, has been continuously developed \cite{hoel2018automated}.   
However, one significant limitation of the learning-based approaches is their valid explanations and causal reasoning for trusting issues \cite{expaisurvey, expai}: the passengers would hardly trust the autonomous vehicles if their decisions cannot be explained. 
Although Explainable Artificial Intelligence has been widely studied \cite{expaisurvey}, it is arguable whether the existing methods are adequate to provide trust in learning-based decision-making for autonomous vehicles \cite{expai}.

Human-like decision making, which can mimic human's decision-making
ability, is a challenging problem as it is difficult to model drivers' 
potential decision-making patterns in the driving process. Game theory,
which can capture the mutual interactions of vehicles and execute
corresponding control actions, has been considered as a promising solution
to the above problem \cite{yu2018human}. In game-theoretic approaches, the
players (vehicles) decide their actions by optimising their profit (here,
safety and speed) in response to the actions of others. 
Many studies developed game-theoretic models in the domain of autonomous driving, e.g., driver behaviour model \cite{albaba2020driver}, lane-changing model \cite{talebpour2015modeling}, and the model of vehicle flows \cite{li2019game}. 
Moreover, several pieces of work investigated game-theoretic approaches to decision-making problems of autonomous vehicles at unsignalised intersections \cite{li2018game,elhenawy2015intersection,wei2018intersection} and roundabouts \cite{banjanovic2016autonomous,tian2018adaptive}.


This paper studies a game-theoretic decision process at roundabouts, which  
are often used to improve traffic safety in urban areas.
According to various studies, the replacement of signalised intersections
by roundabouts reduces injury crashes by 75$\%$
\cite{deluka2018introduction}, and is well-suited to a  low-traffic-volume intersection
\cite{manage2003performance}. From a decision-making point of view, roundabouts
are similar to unsignalised intersections, as they require drivers to
decide when to enter. This decision making depends on the other vehicles 
and the influence of their behaviours. Although roundabouts are
safer than traditional signalised intersections for human drivers, there
are still issues to enforce safety for autonomous vehicles. The inner
island of the roundabout limits the ability of autonomous vehicles to
predict traffic patterns and may lead to traffic collisions. Therefore,
critical decision making is the key to collision-free driving at
roundabouts.

Game-theoretic decision-making approaches at roundabouts can be divided into two groups: machine-learning and game-theoretic ones. Machine-learning approaches feed data (e.g., traffic images) to a machine-learning component and make decisions using classification algorithms \cite{wang2018camera,wang2019multi,garcia2019autonomous}.
Although these approaches are especially suitable for decision making at uncertain roundabout environments, they require massive data sets to train on and are highly susceptible to errors. In contrast, the game-theoretic approach does not require training data and is able to provide a human-like decision by considering the vehicles as players of a game. In \cite{tian2018adaptive}, Tian et al. proposed a cooperative strategy in conflict situations between two autonomous vehicles at a roundabout using non-zero-sum games. Each autonomous vehicle aims to minimise its waiting time by analysing all possible actions and influences of
other vehicles on the game outcome. In \cite{tian2018adaptive} and
\cite{li2018game}, the authors applied k-level games to decision making in various uncontrolled intersection scenarios.


In this paper, we propose a distributed decision-making process for autonomous
vehicles engaging within a roundabout.  
We introduce the notion of \emph{aggressiveness}, which indicates how much each vehicle values speed over safety,
and use it for modelling the behaviours of the vehicles.
At each time step, each vehicle
estimates other vehicles' aggressiveness
based on their observations.
These estimations allow the vehicle to formulate vehicle interactions as a finite perfect-information sequential game.
Based on the Nash equilibrium of this game, the vehicle
predicts other vehicles' behaviours and
makes decisions.
We perform numerical simulations to illustrate the
effectiveness of the proposed process, both for safety (absence of
collisions), and speed (time spent within the roundabout).

\subsection*{Outline}
The rest of this paper is organised as follows. We first set up the scenario and formulate the problem in Section 2. Then, we present the main flow of our decision-making process in Section 3. 
We dedicate Section 4 to describe the cost functions used in the proposed decision-making process in details.
In Section 5, we perform a set of simulations to demonstrate the effectiveness of the proposed approach. Finally, Section 6 concludes the paper.

\subsection*{Notations}
We use the bold font (e.g., $\bm{x}_i$, $\bm{a}_j$) to denote vectors.
We  write $[z(j)]_{j \in N}$ for the vector 
$\begin{bmatrix} z_{j_1} & z_{j_2} & ... & z_{j_n} \end{bmatrix}$,   
where $N = \{j_1 < \ldots < j_n\}$ is a subset of natural numbers. 
We use $[z(j,k)]_{j \in N, k \in M}$, where $N = \{j_1 < \ldots < j_n\}$
and $M = \{k_1 < \ldots < k_m\}$ are subsets of natural numbers,
to denote the vector of vectors $\bigl[[z(j,k)]_{j \in N}\bigr]_{k \in M}$.
When $z_j$ is a parameter (aggressiveness, navigation path, cost function)
of the vehicle $j$, we reserve the notation $\iest{z}{j}{t}$ for its
\emph{estimation by the vehicle $i$ at time $t$}.
Similarly, when $z_j$ is a value (configuration, acceleration) related
to the vehicle $j$, we reserve the notation $\ijpred{z}{t}{\tau}$ for its
\emph{prediction by the vehicle $i$ at time $t$ in $\tau$ time steps into
the future}.
Finally, we reserve the notations $\ijcomp{z}{t}$ and $\ijcomp{z}{t,\tau}$
for intermediate functions that will help us compute the estimation
$\iest{z}{j}{t}$ or the prediction $\ijpred{z}{t}{\tau}$ (in
Sections~\ref{section:nextconfig} and \ref{subsec:accucost}). 

The reason why we use the unusual notations $\langle t \rangle$ and
$\langle t,\tau \rangle$ to express the dependency on the time steps of
those estimations and predictions is to separate this dependency from their
inputs. 
For example, the vehicle dynamic function
$f_j$ of the vehicle $j$ is a function
that takes a configuration $\bm{x}$ and an acceleration $a$  as inputs, and
outputs a new configuration $f_j(\bm{x},a)$. 
With our notations, $\iest{f}{j}{t}$ is the
estimation of the function $f_j$ by the vehicle $i$ at time $t$,
where $\iest{f}{j}{t}$ and $f_j$ are functions of the same type.
Thereby, we write  $\iest{f}{j}{t}(\bm{x},a)$ for the configuration that
the vehicle $j$ would reach in one time step starting from $\bm{x}$ with
acceleration $a$, assuming it follows the path given by the estimation
$\iest{f}{j}{t}$.

\section{Problem formulation}\label{section:model}

This section introduces our problem setting. We first describe 
the traffic scenario in consideration, then the vehicle dynamics model, and 
finally, the goal of this paper.

\subsection{Scenario of interest}\label{section: scenario}

This paper considers the decision making of $n$ autonomous vehicles at a
single-lane $w$-way unsignalised roundabout intersection. 
We assume that the vehicles do not communicate with 
each other, that entrances and exits of the roundabout are right-hand
traffic, and that the traffic flows counter-clockwise within the roundabout. 
A vehicle may use any entrance and 
any exit of the roundabout but is not allowed to drive backward.
Fig.~\ref{fignavex2} illustrates a four-way roundabout. 
We study the decision making based on vehicles within the roundabout and those
approaching the roundabout, but independent of those that have already exited the
roundabout.
 
\subsection{Vehicle configuration setup}\label{section:path}

\begin{figure}
\centering 
{\includegraphics[width=0.25\textwidth]{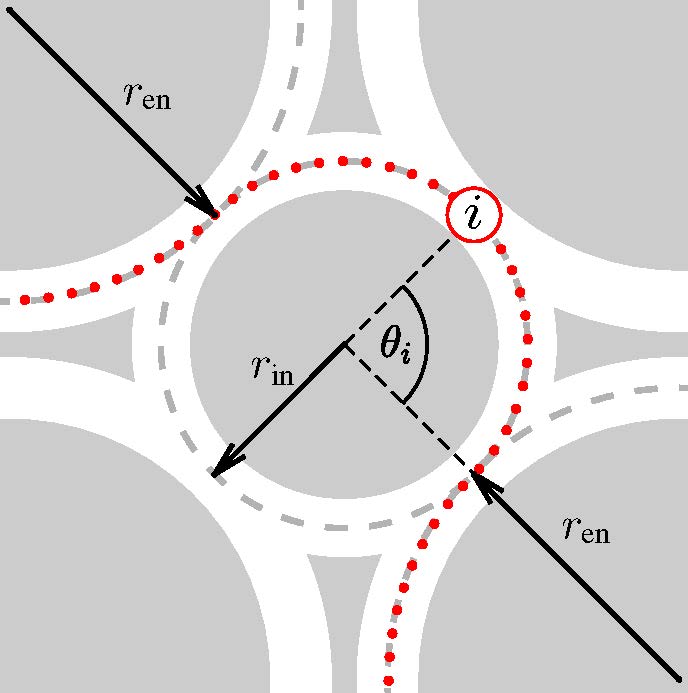}}
 \caption{A four-way roundabout and an example of a navigation path ({\color{red}red} dotted line) of a vehicle $i$
 }
 \label{fignavex2}
\end{figure}

\begin{figure} [t]
\begin{center}
\includegraphics[width=0.75\linewidth]{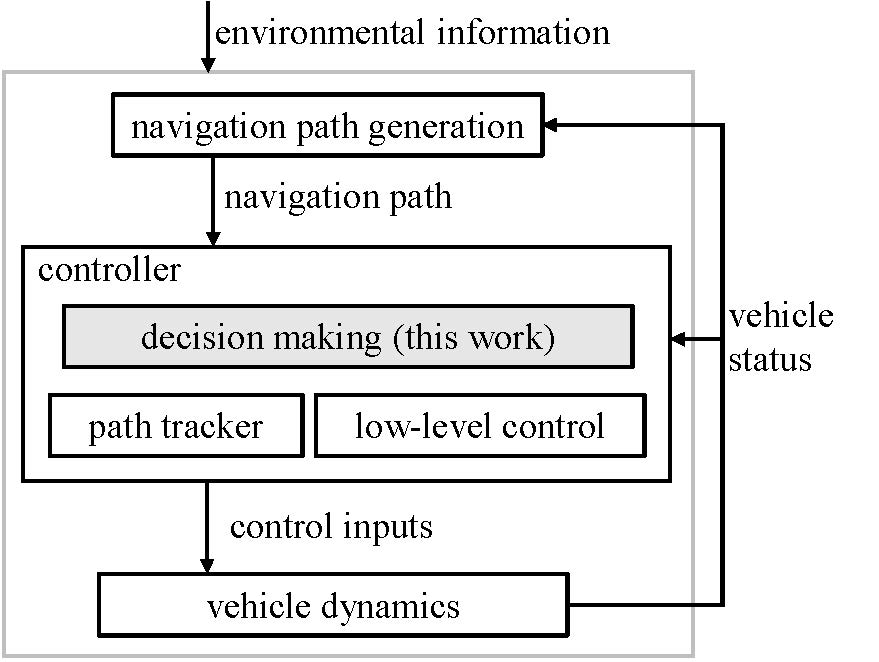}
\end{center} 
\caption{Motion planning architecture~\cite{perez2009, liu2017creating} and the position of this work
\label{fig:position}} 
 \end{figure}

In order to focus on the high-level decision making for autonomous 
vehicles, this paper does not consider the path trackers and the low-level control layer, which 
controls the engine to follow a precomputed navigation path (see Fig.~\ref{fig:position}).
For this reason, we assume that  a vehicle can perfectly follow a given
navigation path.

A navigation path can be arbitrary as long as it is defined by a \emph{vehicle dynamic function} that takes the current configuration and  acceleration as arguments and returns the configuration in the next time step. 
Fig.~\ref{fignavex2} presents an example of a navigation path used in our experiment. In our framework, the configuration of a vehicle $i \in \{1,\ldots, n\}$ at time $t$ is given by:

\begin{equation}
\CONF{i}(t)=
\begin{bmatrix}
r_i(t) & \theta_i(t) & v_i(t) & \MAN{i}(t)
\end{bmatrix}
 \label{eq:configuration}
 \end{equation}
where
$r_i(t)$ and $\theta_i(t)$ indicate the \emph{position} of the vehicle $i$ in polar coordinates, $v_i(t)$ is its \emph{speed} (velocity along the path), and
  $\MAN{i}(t) \in \{\text{``enter'' }, \text{``inside'' },  \text{``exit''}\}$ is its \emph{status} at time step $t$, respectively.
  Then, the time-evolution of the configuration 
  of each vehicle $i$ is represented by 
\begin{equation}
 \CONF{i}(t+1) = \NEXT{i}(\CONF{i}(t), a_i(t)),
 \label{eq:nextstep}
 \end{equation}
where the vehicle dynamic function  $\NEXT{i}$ returns the configuration of the vehicle 
$i$ after one time step, 
 assuming that $a_i(t)$ is a constant 
 acceleration along the path (the derivative of speed)
 between the 
 time steps $t$ and $t+1$.

 For example, in the case where the vehicle is within the roundabout (e.g., the vehicle $i$ in Fig.~\ref{fignavex2}), $\CONF{i}(t+1) = \NEXT{i}(\CONF{i}(t),a_i(t))$ is given by the dynamics as follows:
\begin{align}\label{eq theta}
\begin{split}
r_i(t+1) &=r_\text{in}\\
v_i(t+1) &= v_i(t) + a_i(t)\Delta\\
\theta_i(t+1) &= \theta_i(t) + \frac{v_i(t)}{r_{\text{in}}}\Delta +\frac{1}{2}\frac{a_i(t)}{r_{\text{in}}}\Delta^2\\
s_i(t+1) &= \text{``inside''},
\end{split}
\end{align}
 where $\Delta$ is the duration between two time steps. The cases where the vehicle is entering or exiting are similar.

Each vehicle decides its acceleration $a_i(t)$ at every time step $t$.
To simplify the problem, we further assume that each 
vehicle $i$  chooses an acceleration $a_i(t)$  from a finite set at each time step $t$ to minimise their cost functions. This set of accelerations can theoretically be arbitrary, as long as it is finite. However, the larger this set is, the slower the computation of Nash equilibria will be. We provide concrete values in Section~\ref{section:experiments}.

\subsection{Goal: an efficient distributed decision-making process} 
\label{section:aggressive}

The goal of our paper is to design a process for decision making, which is a distributed algorithm to control the autonomous vehicles.
The proposed decision-making process will be evaluated in 
Section~{\ref{section:experiments}}  to ensure that multiple vehicles can operate simultaneously within a roundabout  and can reach their target safely within an acceptable time. 

 Each vehicle computes its control input 
(namely, its acceleration)
by trying to 
optimise not only its own objective but also its estimations of the objectives of the other vehicles.
We define those objectives using cost functions, which 
specify how bad the situation is, considering two features: safety and 
speed.
The safety feature is small when the distance with other vehicle is large, while the speed feature is small when the vehicle's speed is close to the maximal legal speed. 
The trade-off is defined by the weight vector 
\begin{equation}
\bm{w}_i = [w_i^s, w_i^a], \ w_i^s + w_i^a = 1,
 \label{eq:aggressiveness}
 \end{equation} 
 where $w_i^a \geq 0$ is the weight of the speed feature,  
and $w_i^s = 1-w_i^a \geq 0$ is the weight of the safety feature.  
 This weight vector indicates how much the vehicle $i$ values speed over safety: the higher $w_i^a$ is (and so, the lower $w_i^s$ is), the more aggressive the vehicle will be (and so, the less conservative). 
 Therefore, we call $w_i^a$ the \emph{aggressiveness} of vehicle $i$. 
This value plays a significant role in our decision-making process.

\section{Decision-making process} 
 
In this section, we describe our proposed decision-making process for computing the control input at each time step for each vehicle.
We first introduce an overview in Section~\ref{subsec:outline},
then elaborate each step in detail in Sections~\ref{section:observations}-\ref{section: update}.  
The computation of control inputs, which is the key step of the decision making,
is presented in Section~\ref{section:decision} using the cost functions defined in Section~\ref{section:cost}.

\subsection{Outline of the decision-making process}
\label{subsec:outline}

\begin{figure} [t]
\begin{center}
\includegraphics[scale = 0.5]{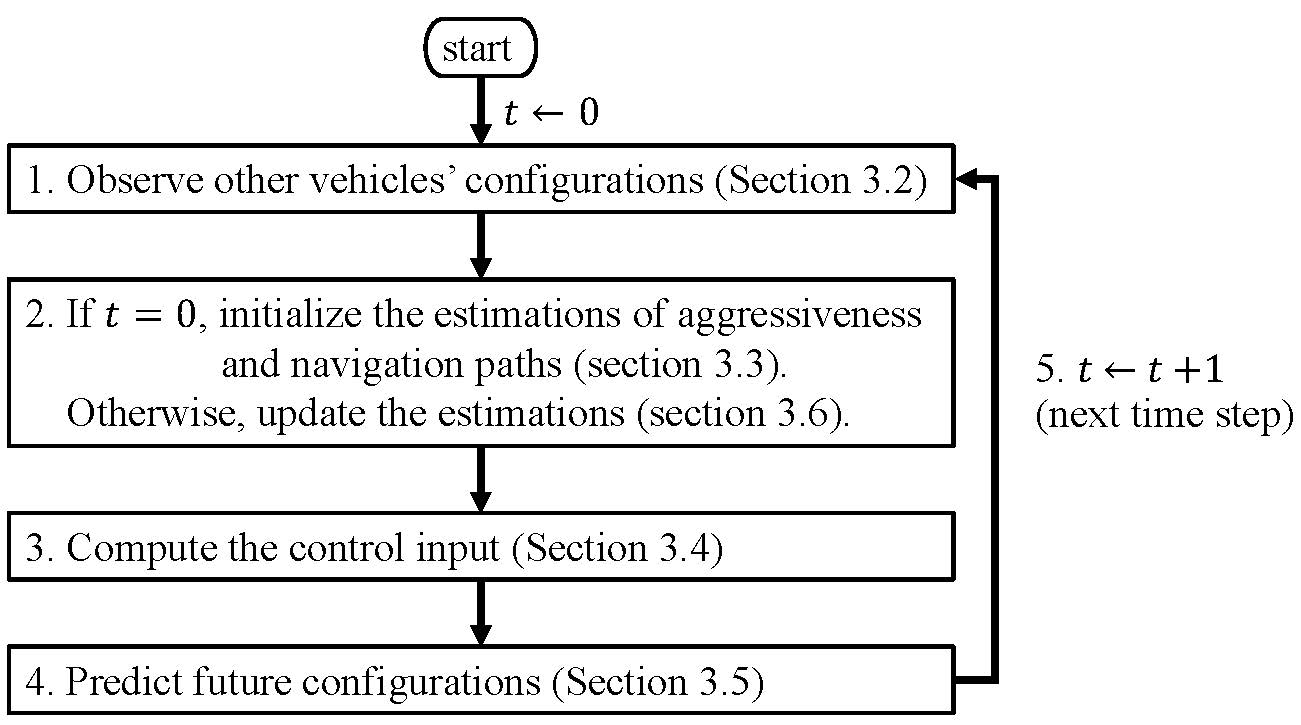}
\end{center} 
\caption{Outline of the decision-making process. Each vehicle repeats the process until it exits the roundabout.  
\label{fig:algo}} 
 \end{figure}

We first introduce an overview of our proposed decision-making process (see also Fig. \ref{fig:algo}). 
Each vehicle $i$ repeats the following steps until it exits the roundabout.
\begin{enumerate}
	\item The vehicle $i$ observes the current configurations of its neighbours, which are the vehicles nearby (Section~\ref{section:observations}).
	\item The vehicle $i$ initialises the estimations of navigation paths and aggressiveness of the neighbours (Section~\ref{section: initialisation}) or updates those estimations by comparing the newly observed configurations with its predictions in the previous time step (Section~\ref{section: update}).
	\item Based on the estimations, the vehicle $i$ computes its control input (acceleration) and predicts  accelerations of its neighbours (Section~\ref{section:decision}).
	\item The vehicle $i$ predicts its neighbours' future configurations up to a finite time horizon, based on the estimated navigation path and aggressiveness (Section~\ref{section:nextconfig}).
	\item Then, the vehicle $i$ operates by using the computed control input and repeats the process for the next time step.  
\end{enumerate} 

\subsection{Observations}\label{section:observations}
In the roundabout, each vehicle $i$ may not be able to observe all other vehicles, but only the vehicles nearby. 
Let $\NEI_i(t)\subseteq \{ 1, \ldots, m \}$ be the set of vehicles that are being 
observed by the vehicle $i$, called the \emph{neighbours} of $i$, at time step $t$. 
We assume that $i \in \NEI_i(t)$ as $i$ can always observe itself. 
In our simulations, $\NEI_i(t)$ consists of $i$ itself, 
the closest two vehicles in front, and the closest one behind, 
if the distance is smaller than a fixed value $D$.
The \emph{observation} of $i$ at time $t$ is the 
collection of all the individual configurations $\CONF{j}(t)$ for 
$j \in \NEI_i(t)$.
We use $[\CONF{j}(t)]_{j \in \NEI_i(t)}$ to denote the vector of observed configurations, also called \emph{observations}, of all the neighbours of $i$.

\subsection{Initialisation of the estimations of aggressiveness and navigation paths of neighbours}\label{section: initialisation} 
Initially (at time step $t = 0$),
the vehicle $i$ estimates the aggressiveness $w_j^a$ of any vehicle $j\in \NEI_i(0)$ to be $\iest{w^a}{j}{0} = 0.5$. 

For the initial navigation paths, we cannot assume that the vehicle $i$ knows the vehicle dynamic function $\NEXT{j}$ as $i$ does not know which path the vehicle $j$ will follow. 
For example, $i$ does not know which exit $j$ will use.  
Therefore, initially, the vehicle $i$ estimates a vehicle dynamic function $\iest{f}{j}{0}$ using the initial configuration $\CONF{j}(0)$ of the vehicle $j$. 
Intuitively, if the vehicle $i$ detects that the vehicle $j$ is steering out, then it knows that $j$ will use the next exit. Otherwise, $i$ computes the navigation path as if the vehicle $j$ stayed within the roundabout indefinitely, possibly turning around several times. 
We always have $\iest{f}{i}{t} = \NEXT{i}$ at any time $t$ as $i$ knows its own navigation path.

\subsection{Control inputs and predictions, as Nash equilibria} \label{section:decision}

We formulate a non-cooperative sequential game played between the vehicles to decide the acceleration at each time step (see
Appendix for details).
Each vehicle $i$ determines $a_i(t)$  at time $t$
using the Nash equilibrium of a sequential game 
\begin{equation*}
G_i(t)=(\NEI_i(t), \Sigma, [\iest{K}{j}{t}]_{j\in\NEI_i(t)}, \prec),
\end{equation*} 
defined as follows.
\begin{enumerate}
 	\item The players are the neighbours, i.e., elements of $\NEI_i(t)$. 
	\item A strategy of each vehicle $j \in \NEI_i(t)$ is a vector of accelerations $\sigma_j = [a_j(t,\tau)]_{0\leq \tau < h} \in \Sigma$ for the time-step horizon from $\tau = 0$ to $\tau = h-1$. Namely, $a_j(t,\tau)$ is an acceleration of the vehicle $j$ at time step $t+\tau$. Recall that $a_j(t,\tau)$ is selected from a finite set of accelerations (see Section~\ref{section:path}). Therefore, the set $\Sigma$ of all possible strategies is also a finite set.
	\item Let $[\sigma_j]_{j\in\NEI_i(t)}$, where $\sigma_j\in \Sigma$, 
	 be a vector of strategies for all vehicles in $\NEI_i(t)$.
	The accumulated cost of vehicle $j \in \NEI_i(t)$ for the time horizon from $\tau = 0$ to $\tau = h-1$ estimated by the vehicle $i$ at time $t$ is 
	\begin{align}
	\begin{split}
		\iest{K}{j}{t}([\sigma_k&]_{k\in\NEI_i(t)}) = \\
		&\ijcomp{K}{t}(\NEI_i(t), \iest{w^a}{j}{t},[\sigma_k]_{k\in\NEI_i(t)} ),
		\end{split}
		\label{eq:gamecost}
	\end{align}
where the definition of  the function $\ijcomp{K}{t}$ is given by Eq.~\eqref{eq:hcost} in Section~\ref{section:cost}.
	The objective of each vehicle playing this game is to minimise its accumulated cost.
	\item 
	The order $\prec$ of the players in the sequential game is according to the estimated aggressiveness: if $\iest{w^a}{j}{t} < \iest{w^a}{k}{t}$, then $k$ makes the decision before $j$. In other words, the more aggressive a vehicle is, the more priority it will have to choose its control input (i.e., its acceleration).
\end{enumerate}

Let $[\ijpred{a}{t}{\tau}]_{j\in\NEI_i(t), \, 0\leq \tau < h}$ 
be a Nash equilibrium of this game, which always exists and is computable because the game is a finite perfect-information game~\cite{O2009}.  
Then, $\ijpred{a}{t}{\tau}$ is the acceleration of the vehicle $j$ at time $t + \tau$ that vehicle $i$ predicts at time $t$.
We compute this Nash equilibrium  using backward induction (see Appendix for details). The theoretical complexity of this computation is $O(h\cdot\lvert\NEI_i\tbrac{t}\rvert\cdot\lvert\Sigma\rvert^{\NEI_i\tbrac{t}})$. Observe that this complexity is exponential in the number of players, which explains why we restrict these games to be played only among neighbours.

Finally, we use $\ipred{a}{t}{0}$ as the control input of vehicle $i$ at time step $t$.

\subsection{Prediction of future configurations}\label{section:nextconfig}


For each vehicle $j \in \NEI_i(t)$, let  
$\ACC_j = [a_{j}(t,\tau)]_{0 \leq \tau < h}$ 
be a given vector of accelerations of $j$.
For example, those accelerations can be a Nash equilibrium computed in Section~\ref{section:decision}, namely, $\ACC_j = [\ijpred{a}{t}{\tau}]_{0 \leq \tau < h}$.

Let $\ijcomp{\bm{x}}{t,\tau}(\ACC_j)$ be the configuration of the vehicle $j$ that would be reached at time step $t+\tau$, based on the observation $\CONF{j}(t)$, the estimations $\iest{f}{j}{t}$, and the vector of accelerations $\ACC_j$.   
These configurations are computed by induction as follows. 

\begin{itemize}
	\item The configuration at time step $t$ is the observed configuration
		\begin{equation}
			\ijcomp{\bm{x}}{t,0}(\ACC_j) = \CONF{j}(t).
			\label{eq:initialprediction}
		\end{equation}
	\item The configuration at time step $t+\tau+1$ is computed by assuming that the vehicle $j$ will follow the estimated navigation path computed using $\iest{f}{j}{t}$, from the configuration $\ijcomp{\bm{x}}{ t,\tau}(\ACC_j)$, with acceleration $a_{j}(t,\tau)$:
		\begin{equation}
			\ijcomp{\bm{x}}{t,\tau+1}(\ACC_j) = \iest{f}{j}{t}\Bigl(\ijcomp{\bm{x}}{t,\tau}(\ACC_j), a_{j}(t,\tau)\Bigr).
			\label{eq:inductiveprediction}
		\end{equation}
\end{itemize}

In particular, we define 
the predicted configuration of the vehicle $j$ at time $t + \tau$ which vehicle $i$ predicts at time $t$ as
        \begin{equation}
            \ijpred{\bm{x}}{t}{\tau} = \ijcomp{\bm{x}}{ t,\tau}([\ijpred{a}{t}{\tau}]_{0 \leq \tau < h}).
        \end{equation}

\subsection{Update of the estimations of aggressiveness and navigation paths} \label{section: update}

At each time step $t$, let 
 $U_i(t) \subseteq \NEI_i(t)\setminus \{i\}$ be the set of vehicles  
such that
the distance between the observed configuration $\CONF{j}(t+1)$ and the predicted configuration $\ijpred{\bm{x}}{t}{1}$
is bigger than a fixed value. 
Since the predictions of the vehicle $i$ are not precise, $i$ needs to update its estimations of aggressiveness and navigation path for each vehicle $j \in U_i(t)$.

For the aggressiveness, the vehicle $i$ updates the estimation $\iest{w^a}{j}{t}$ to $\iest{w^a}{j}{t+1}$, which describes the behaviour of the vehicle $j$ more accurately. 
The vehicle $i$ considers each value $w^a$ in a given fixed finite set $\mathcal{W}$ 
and constructs
a game $G_{i,j}(t,w^a)$ in the same way as $G_i(t)$ in Section \ref{section:decision}.  
Concretely, we construct $G_{i,j}(t,w^a)$ as follows. 
\begin{enumerate}
    \item The set of players is $\{i,j\}$.
    \item The strategies are the same as in Section~\ref{section:decision}.
    \item Given strategies $\sigma_i$ and $\sigma_j$ for the vehicles $i$ and $j$ respectively, their accumulated cost for every vehicle $l\in\{i,j\}$ in this game is  
    \begin{equation} 
		\icomp{K}{l}{t}(\{i,j\}, w^a,[\sigma_k]_{k\in\{i,j\}}),
		\label{eq:updatecost}
	\end{equation}
	where $\icomp{K}{l}{t}$ is the same function as in Section~\ref{section:decision}, and which will be described in Section~\ref{section:cost}.
	\item The order $\prec$ of players is defined in the same way as in Section~\ref{section:decision}.
\end{enumerate}
Let $[\kcomp{a}{k}{t,\tau}(w^a)]_{k\in\{i,j\}, \, 0\leq \tau < h}$ denotes the Nash equilibrium of this new game $G_{i,j}(t,w^a)$.
Then, the vehicle $i$ computes a new estimated aggressiveness $\iest{w^a}{j}{t+1}$ of the vehicle $j$ that fits the observed acceleration $a_j(t+1)$ the most closely, i.e.,
\begin{equation}
 \iest{w^a}{j}{t+1} =  \displaystyle\arg\min_{w^a\in\mathcal{W}} \,\lvert \kcomp{a}{j}{t,1}(w^a) - a_j(t+1) \rvert
    \label{Equation:Estimator1}
\end{equation}

Otherwise, if the predicted configuration is close enough to the observed configuration ($j \notin U_i(t)$), then $\iest{w^a}{j}{t+1} = \iest{w^a}{j}{t}$.
The estimated vehicle dynamic function $\iest{f}{j}{t}$ is updated in the same way as in Section~\ref{section: initialisation}.

\subsection{Dealing with deadlocks} 
In this section, we consider the situation where all vehicles in $\NEI_i(t)$ are stopped, that is, have zero speed. This situation is particularly critical, as it may induce a deadlock: every vehicle is waiting for the other vehicles to move.

In this case, we want the vehicle $i$ to make a move, as long as it is not in a critical situation, i.e., the situation when $i$ is waiting to enter the roundabout but observes that a vehicle is already inside the roundabout. 
If the vehicle $i$ is not in such a critical situation, we enforce it to make a move by setting the acceleration $a_i(t)$ to $10\,ms^{-2}$ with probability $0.5$.

\section{Cost functions in the sequential game}\label{section:cost}

In this section, we introduce the cost function to be minimised in the sequential game in Section~\ref{section:decision}. For the entire Section \ref{section:cost}, we 
use the following notations.
\begin{itemize}
\item $\bm{x}_i$ is a configuration -- either observed or predicted -- of the vehicle $i$.
\item $\NEI$ is a subset of vehicles. This subset could be the set $\NEI_i(t)$ of neighbours or the set $\{i,j\}$ in Section~\ref{section: update}.
\item $\aggcost$ is a value of the aggressiveness of vehicle $j$. This value $\aggcost$ can either be the estimated aggressiveness $\iest{w^a}{j}{t}$ or the value $w^a$ in Section~\ref{section: update}.
\end{itemize}

 \subsection{Accumulated cost function}
 \label{subsec:accucost} 
We construct the accumulated cost using the receding horizon control approach~\cite{K2005}, based on the predicted future up to a horizon time step $h < \infty$. 
Recall that we use this accumulated cost to determine the control inputs of each vehicle $i$ in Section~\ref{section:decision}, and to update the estimation of the aggressiveness of other vehicles in Section~\ref{section: update}.

As introduced in Section~\ref{section:decision}, a strategy of each vehicle $j \in \NEI$ is a vector of accelerations $\sigma_j = [a_j(t,\tau)]_{0\leq \tau < h}$ for the time-step horizon from $\tau = 0$ to $\tau = h-1$.
The accumulated cost function of the vehicle $j$, computed by the vehicle $i$ at time step $t$, based on the aggressiveness value $\aggcost$ is

\begin{align}
\begin{split}
\ijcomp{K}{t}&(\NEI, \aggcost ,[\sigma_k]_{k\in\NEI} ) \\
&= \sum_{\tau=0}^{h-1} \lambda^\tau \cdot
 \SCOST_j([\icomp{\bm{x}}{k}{t,\tau}(\sigma_k)]_{k\in\NEI},\NEI, \aggcost),
 \end{split}
\label{eq:hcost}
\end{align}
where $\lambda\in (0,1)$ is a fixed discount factor, $[\icomp{\bm{x}}{k}{t,\tau}(\sigma_k)]_{k\in\NEI}$ is the predicted configuration at time step $t + \tau$ that was defined in Section~\ref{section:nextconfig}, and $\SCOST_j$ is the time-step cost function that will be defined in Section~\ref{section:step cost}.

Let us remark that the vehicle $i$ uses its own observations to compute the accumulated cost of the vehicle $j$. 
Specifically, in Eq.~\eqref{eq:gamecost}, $i$ computes the  accumulated cost of $j$ using its own neighbours $\NEI_i(t)$. 
In the case where $j$ is the nearest vehicle in front of $i$, we have two situations depending on whether or not $i$ can observe its second nearest vehicle in front. If it can, then this vehicle will be in $\NEI_i(t)$ and will be considered as the nearest vehicle in front of $j$ in the computation of $\ijcomp{K}{t}$. If it cannot, this means that the second nearest vehicle is too far from $i$, so that $i$ will compute $\ijcomp{K}{t}$ as if there were no vehicle in front of $j$.

\subsection{Cost at each time step}\label{section:step cost}
We introduce the cost of a vehicle at each time step that we call \emph{time-step cost} and use it for computing the accumulated cost in Eq.~\eqref{eq:hcost}.
The time-step cost function of the vehicle $i$ is given by
\begin{align}\begin{split}
&\SCOST_i([\bm{x}_j]_{j \in \NEI},\NEI, \aggcost)\\
&~= (1 - \aggcost) \phi_i^\text{safe}([\bm{x}_j]_{j \in \NEI}, \NEI)
+ \aggcost \phi_i^\text{speed}([\bm{x}_j]_{j \in \NEI}, \NEI),
\end{split}\label{eq:stepcost}
\end{align}
where 
$[\bm{x}_j]_{j \in \NEI}$
is a given vector of configurations of vehicles in $\NEI$, 
$\phi_i^\text{safe}$ and $\phi_i^\text{speed}$ are safety and speed features
defined in Sections~\ref{section:safety} and \ref{section:speed}, respectively.

\subsubsection{Safety feature}\label{section:safety}
In order to evaluate the safety feature, each vehicle $i$ considers the nearest vehicle in front of and the nearest one behind within a given distance $D$ along the navigation path (if they exist). 
Concretely, we consider the pair of vehicles $(i^+,i^-) \in \NEI \times \NEI$ such that
\begin{align*}
i^+=\displaystyle\arg\!\min_ {f} &\{ \theta_f-\theta_i \mid \\
& f \in \NEI\setminus \{i\} , 0 \leq \theta_f-\theta_i  \leq \pi,   d(f,i) < D\}, \\
i^-= \displaystyle\arg\!\min_ {b } &\{ \theta_i-\theta_b\mid \\
& b \in \NEI\setminus \{i\} , 0< \theta_i-\theta_b <\pi, d(b,i) < D\},
\end{align*}
where $\theta_j$ is the angular position of vehicle $j$ with respect to the centre of the roundabout and $d(i,j)$ is the distance between $i$ and $j$ measured along the navigation path. Namely, $i^+$ (\emph{resp.} $i^-$) is the nearest vehicles in front of (\emph{resp.} behind) the vehicle $i$ within the distance $D$.

Then, we define the safety feature $\phi_i^\text{safe}$ as
\begin{align}
\begin{split}
\phi_i^\text{safe}&([\bm{x}_j]_{j \in \NEI}, \NEI)= \\
&\max(\phi_i^\text{front}([\bm{x}_j]_{j \in \NEI}, \NEI),\phi_i^\text{back}([\bm{x}_j]_{j \in \NEI}, \NEI)).
\end{split}
\label{eq:stepcostfeatures}
\end{align}
The feature $\phi_i^\text{front}$ is given by  
\begin{align}
\begin{split}
 \phi_i^\text{front}\Big(\Big[\bm{x}_j& = 
 \begin{bmatrix}
 r_j  & \theta_j & v_j& \MAN{j} 
 \end{bmatrix}\Big]_{j \in \NEI} , \NEI\Big) \\
 =
&\begin{cases} 
	0 	&\mbox{if no }  i^+ \mbox{ exist},\\
	C_{\text{ins}}\cdot(D - d(i^+,i))^2
		&\mbox{if }\MAN{i}=\text{inside} \text{ and }\\
		&\qquad\MAN{i^+}=\text{enter},\\
	C\cdot(D - d(i^+,i))^2 
		&\mbox{if }\MAN{i}=\text{enter} \text{ and } \\
		\qquad + \beta(d(i^+,i) ,D_{\text{en}})
		& \qquad \MAN{i^+}=\text{inside},\\
	C\cdot(D - d(i^+,i))^2 &  \mbox{otherwise}\\
		\qquad + \beta(d(i^+,i) ,D_{\text{c}}),
\end{cases}
\end{split}
\label{eq:bfcost}
\end{align}
with
\begin{equation}
\beta(\mathit{dist}, \mathit{threshold}) =
	\begin{cases} 
	E_\infty & \mbox{if } \mathit{dist} \leq \mathit{threshold},\\
	0 & \mbox{otherwise}. 
	\end{cases}
	 \label{eq:beta}
\end{equation}
where 
$C_\text{ins} < C$, $E_\infty$, $D_\text{c} < D_\text{en}$ are given positive constants. 
In Eqs.~\eqref{eq:bfcost}, $\phi_i^\text{front}$ is the cost induced by the nearest vehicle in front of the vehicle $i$, i.e., the vehicle $i^+$. All the cases depend on the statuses of the vehicle $i^+$ and the vehicle $i$. The equations in \eqref{eq:bfcost} reflect the fact that, when $i$ is inside the roundabout (the second case), it does not have to pay much attention to the vehicles that are not yet entered, as $i$ has priority. On the other hand, $i$ has to be extra careful when it is entering the roundabout, as it does not have priority (the third case). The intuition of the equations in \eqref{eq:beta} 
is that a vehicle has to be extra careful when it is close to other vehicles, as the value of $E_\infty$ is large. 

We define the feature $\phi_i^\text{back}$ in the same way as $\phi_i^\text{front}$,
by changing $i^+$ in the equations in \eqref{eq:bfcost} to $i^-$.

\subsubsection{Speed feature}\label{section:speed}
Let $v_\mathit{l}$ be the speed limit of the road. 
The speed feature is given by
\begin{align}
\begin{split}
\phi_i^\text{speed}&\Big(\Big[\bm{x}_j = 
 \begin{bmatrix}
 r_j  & \theta_j & v_j& \MAN{j} 
 \end{bmatrix}\Big]_{j \in \NEI}, \NEI\Big)  \\
&= \begin{cases}     
C_\text{en} \cdot( v_\mathit{l} - v_i)^2 &\mbox{if } v_\mathit{l} \geq v \text{ and } \MAN{i} = \text{enter}\\
C_\text{in} \cdot( v_\mathit{l} - v_i)^2 &\mbox{if } v_\mathit{l} \geq v  \text{ and } \MAN{i} \neq \text{enter}\\
C_\text{o}\cdot(v_\mathit{l} - v_i)^2 &\mbox{otherwise}
\end{cases}
\end{split}
\label{eq:velocost}
\end{align}
where $C_\text{in}$, $C_\text{en}$ (resp. $C_\text{o}$) are constant positive coefficients for the cases that $v_{i}$ is under (resp. over) the speed limit. The intention is that $C_\text{o}$ is much bigger than $C_\text{in}$ and $C_\text{en}$ because we cannot allow a vehicle to exceed the speed limit. 
 

\section{Experimental study}\label{section:experiments}


In this section, we conduct an experimental study to examine the performance of our proposed decision-making process. First, in Section~\ref{Subsection:Experiement.Setup}, we introduce the configuration setting, including the roundabout scenarios and the parameters of the involved vehicles.   
Then, in Section \ref{Subsection:Experiment.ResultsandAnalysis}, we present the numerical simulation results and analyse the performance of the decision making. Specifically, the following issues are addressed and explored.  

\begin{enumerate}
\item \textbf{General performance.}  Can the proposed decision-making process manage the roundabout traffic safely and efficiently?\label{Experiment:RQ1}

\item \textbf{Factors that influence the performance.} Which factors can influence the performance of decision making? \label{Experiment:RQ2}

\item \textbf{Interaction between vehicles.} How do the vehicles interact with each other during the decision-making process?\label{Experiment:RQ3}
\end{enumerate}

We explore the general performance in Section~\ref{Subsubsection:Experiment.ResultsandAnalysis.GeneralPerformance} and present an overall evaluation of the proposed decision-making process.  
Then, in Section \ref{Subsubsection:Experiment.Analysis.Factors}, we explore some factors that influence the vehicles' performance within the roundabout traffic,
in order to explore some potential directions to improve our work
and to obtain more instructions for future applications.  
Finally, in Section \ref{Subsubsection:Experiments.Analysis.Details}, we examine the interaction among the vehicles during the decision-making process and
the accuracy of predicted accelerations based on Nash equilibria in Section~\ref{section:decision}. 
Our results confirm that the vehicles make rational decisions using the purposed  decision-making process.

\subsection{Simulation setup}
\label{Subsection:Experiement.Setup}
\begin{figure}
\centering    
 \subfigure[]
 {\label{fignav}
  \includegraphics[width=5.5cm]{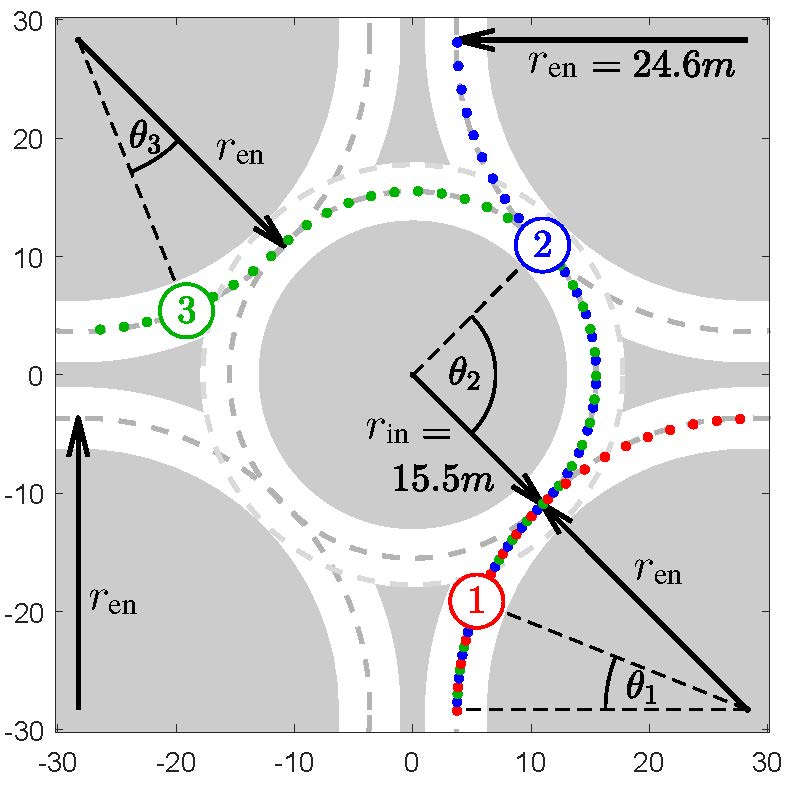} 
  }\\
  \subfigure[]
  {\label{figinipos}
  \includegraphics[width=5.5cm]{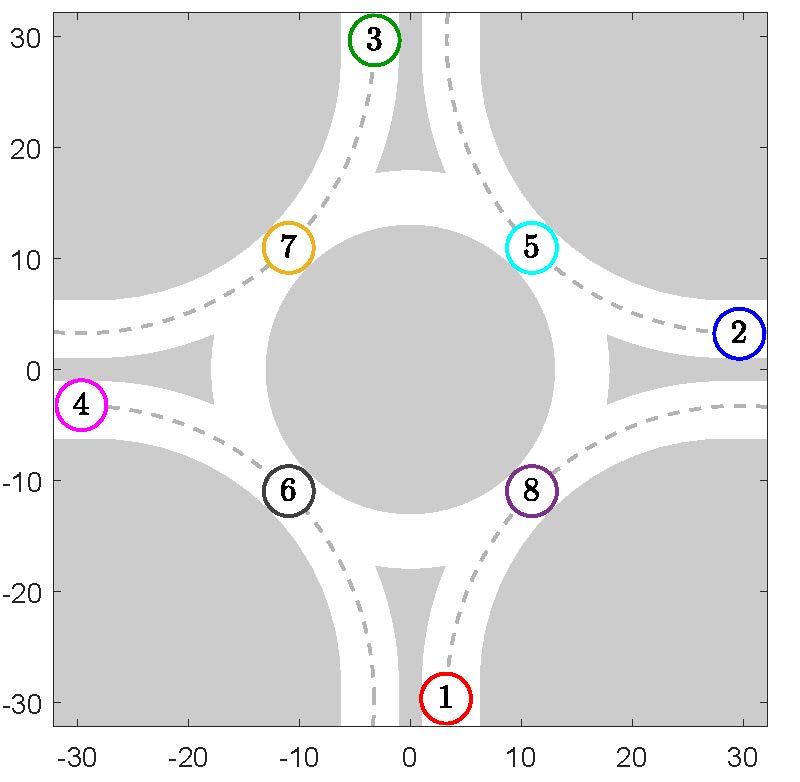}}
   \caption{(a) The roundabout and navigation paths constructed from arcs of five circles: one with radius $r_\text{in}$
 and four with radius $r_\text{en}$.
 Supposing that the vehicles enter from below,
 we consider three types of navigation paths: 
to turn right ({\color{red} red} dotted line), 
to go straight ({\color{blue} blue} dotted line),
and to turn left ({\color{green} green} dotted line).
$\theta_1$, $\theta_2$
and  $\theta_3$ are $0.38$, $\pi/2$, and $0.40$ radians, respectively
(b) The initial positions of the vehicles.
}
\label{fig nav inipos}
\end{figure} 

The roundabout and the navigation paths used in the simulation shown in Fig.~\ref{fignav} 
are designed based on practical situations \cite{anjana2014development}.
All vehicles 
drive on the right-hand
side of the road
and enter the roundabout using one of the four entrances. 
The road occupancy of a vehicle is modelled as a circle with diameter $4.5m$.

The vehicle dynamics is as in Eq.~\eqref{eq theta} with time interval $\Delta = 0.25s$ (following \cite{li18} and \cite{tian18}). 
The status of a vehicle entering (\emph{resp.} exiting) the roundabout changes from ``enter'' to ``inside''  (\emph{resp.} ``inside'' to ``exit'')
when the distance between its centre and the centre of the roundabout is at least (\emph{resp.} greater than) $r_\text{in}+4.5m$, i.e., when it can (\emph{resp.} cannot) possibly collides with other vehicles inside the roundabout. 
In Fig.~\ref{fignav}, vehicles $1$ and $3$  are at the status-changing positions. 

We perform simulations by considering four to eight vehicles, whose types of navigation paths are determined randomly.
The initial positions of the vehicles are shown in Fig.~\ref{figinipos}.
The initial speeds  
are randomly selected from $[0, v_l]$, where $v_l=11ms^{-1}$  is the speed limit of the road~\cite{fha}.
The aggressiveness $w_i^a$ of each vehicle $i$ is randomly selected from $\{0.2, 0.3, \ldots, 0.8\}$.
Table~\ref{table:parameters} presents the setting of parameters. 
In addition, we use the following vectors of acceleration/deceleration sequences for the strategies of the sequential game (the strategies in $\Sigma$, see Section~\ref{section:decision})
with a time horizon $h = 4$ (the unit is $m/s^2$).
        \begin{itemize}
            \item  
            $\begin{bmatrix}
            -50 & 0 & 0 & 0 & 0
            \end{bmatrix}$
            for a strong deceleration,
            \item $\begin{bmatrix}
            -10 & 0 & 0 & 0 & 0
            \end{bmatrix}$ for a small deceleration,
            \item $\begin{bmatrix}
            0 & 0 & 0 & 0 & 0
            \end{bmatrix}$ for no acceleration,
            \item $\begin{bmatrix}
            10 & 0 & 0 & 0 & 0
            \end{bmatrix}$ for a small acceleration, 
            \item $\begin{bmatrix}
            30 & 0 & 0 & 0 & 0
            \end{bmatrix}$ for a strong acceleration.
        \end{itemize}

\begin{table} [t!]
\processtable{Parameters used in the  simulations.\label{table:parameters}}
{\begin{tabular}{@{\extracolsep{\fill}}cccc}\toprule
$\lambda = 0.8$ & $E_\infty = \text{max}\_\text{int}$ & $C_\text{in} =10$ &
 $D =30m$ \\ 
 $D_\text{en} =10m$ & $C_\text{o} =10^3$ &
 $C_\text{ins} = 1$ & $D_\text{c} = 6m$ \\ 
 $v_l =11 m/s$ &
  $C = 10$ & $C_\text{en} =1$  & ~\\\botrule
\end{tabular}}{}
\end{table}

To update the estimation of the aggressiveness (see Section \ref{section: update}),
we use  $\mathcal{W} = \{0.1, \ldots, 0.9\}$, although the actual aggressiveness $w^a_j$ of each vehicle $j$ is initialised within the range $\{0.2, 0.3, \ldots, 0.8\}$.
We allow a vehicle $i$ to update an estimated aggressiveness $\iest{w^a}{j}{t}$ to be $0.1$ and $0.9$ (see Section~\ref{section: update}) so that $i$ can determine whether $\iest{w^a}{i}{t}<\iest{w^a}{j}{t}$ or $\iest{w^a}{i}{t}>\iest{w^a}{j}{t}$. Therefore, $i$ can decide the order of the players for the sequential game in Section~\ref{section:decision}.

 All the programs were coded and run using MATLAB 2018a and MATLAB 2018b.

\subsection{Results and analysis}
\label{Subsection:Experiment.ResultsandAnalysis}

\subsubsection{General performance}
\label{Subsubsection:Experiment.ResultsandAnalysis.GeneralPerformance}

\begin{table} [t!]
\processtable{Summary of simulation results. The columns present the numbers of vehicles, the collision rate, the average minimal distances, and the average time spent at the intersection, respectively. We performed $1000$ simulations for each case, i.e., we perform $5000$ simulations in total. \label{Table:ExperimentalResult}}
{\begin{tabular}{@{\extracolsep{\fill}}cccc}\toprule
\textbf{No. of} & \textbf{Collision} & \textbf{Avg. Minimal} & \textbf{Avg. Mission}\\
\textbf{Vehicles} & \textbf{Rate(\%)} & \textbf{Distances (m)} & \textbf{Time (s)}\\
\midrule
4 & 0 & 14.49 & 10.4 \\
5 & 0 & 9.81 & 12.1 \\
6 & 0 & 8.94 & 13.3 \\
7 & 0 & 8.90 & 14.4 \\

8 & 0 & 8.93 & 15.1 \\
\botrule
\end{tabular}}{}
\end{table}

Table \ref{Table:ExperimentalResult}
summarises the overall performance of our proposed decision-making process.
Columns $1$ to $4$ respectively represent the number of vehicles involved in each simulation,
the percentage of simulations in which a collision occurs,
the average value of the \emph{minimal distance} between two vehicles during each simulation,
and the average \emph{mission time}: the time spent at the roundabout. 

We first consider the safety objective.
For all simulations, even for the $8$-vehicle case,
no collision was detected.
Then, we consider the time efficiency.
In the $4$-vehicle case,  
the vehicles spend approximately $10.4s$ at the intersection.  
In relatively heavy traffic involving eight vehicles, the vehicles spend $15.1s$ on average. 
These results present the effectiveness of the proposed decision-making process.

\subsubsection{Factors that influence the performance}
\label{Subsubsection:Experiment.Analysis.Factors}
We examine the simulation results and observe how the performance changes when increasing the number of vehicles.  
When the number of vehicles is $4$, the average minimal distance is $14.49m$ and the 
average mission time is $10.4s$.
When the number of vehicles reaches $8$, the 
average minimal distance decreases to $8.93m$ and the mission time increases to $15.1s$.
These results indicate that, as more vehicles are involved, the minimal distance decreases while the mission time increases. 

However, 
the results also  
show the rationality of our decision making.  
Although the minimal distance decreases along with the increase in the number of vehicles, the degree of this decrease is not significant. 
From the
$6$ to the $8$-vehicle case, the average minimal distance
changes only slightly. 
This result shows that the vehicles become more cautious as the traffic condition becomes complex, in order to ensure the safety objective.
Also, the increase of the mission time itself, from $10.4s$ in the $4$-vehicle case to 
$15.1s$ in the $8$-vehicle case, is also moderate. 

\begin{figure}
\centering    
 \subfigure[]
 {\label{Subfigure:Experiment.Factors.Aggresive.Dis}
  \includegraphics[width=5.5cm]{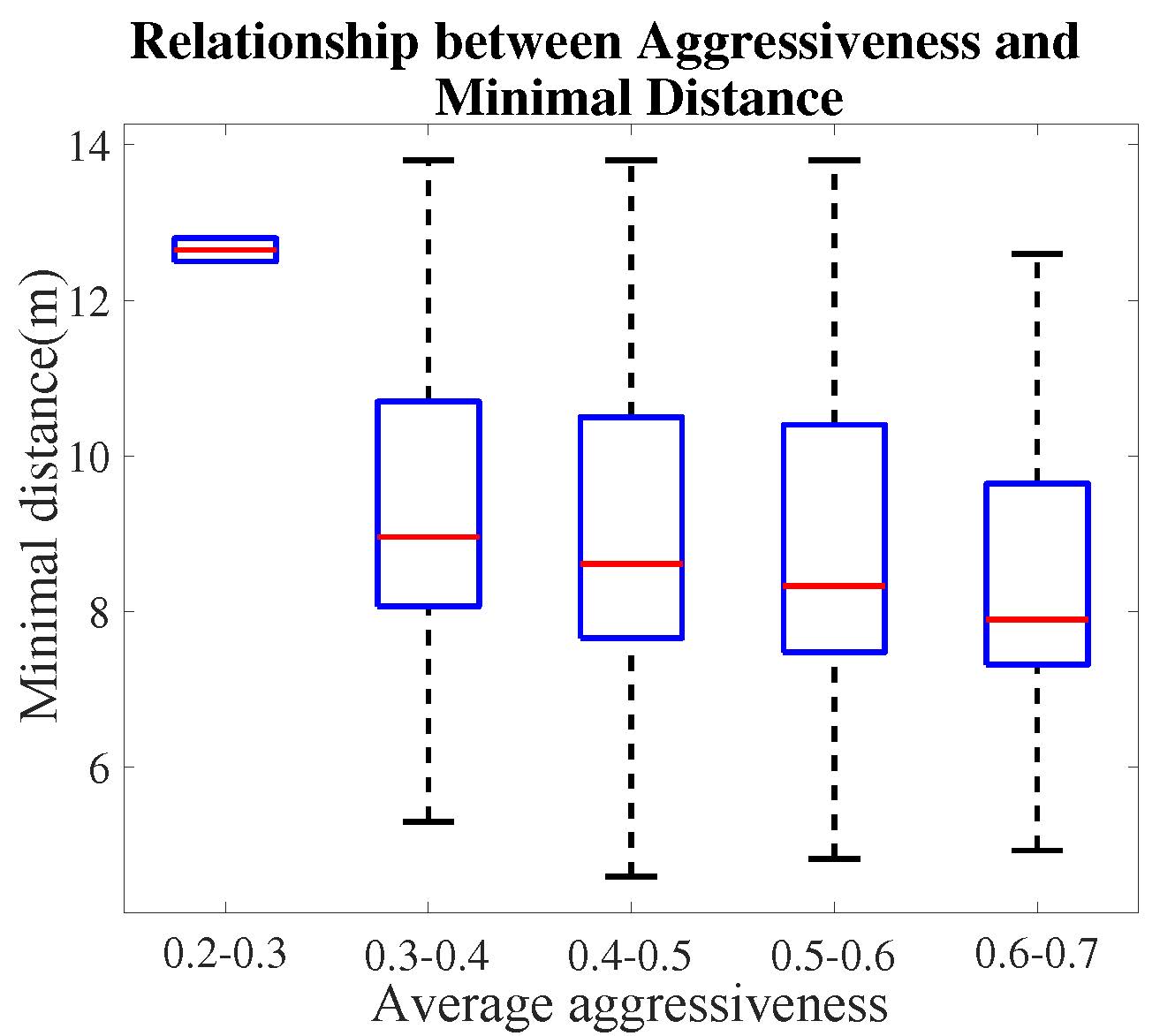} 
  }\\
  \subfigure[]
  {\label{Subfigure:Experiment.Factors.Aggresive.Step}
  \includegraphics[width=5.5cm]{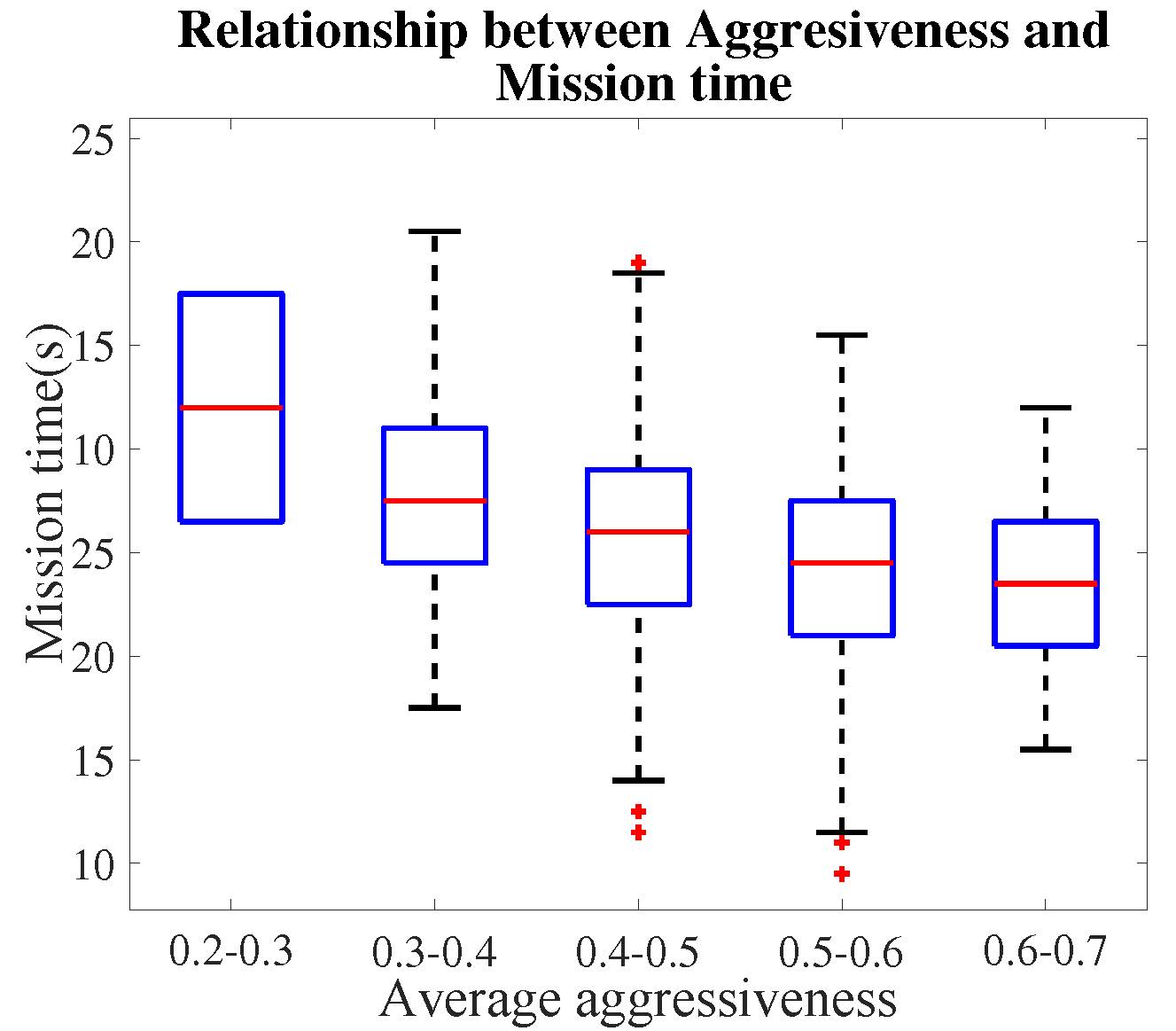}}
   \caption{(a) The box-and-whisker plot describing the relationship between the minimal distances and the average aggressiveness.
   (b) The box-and-whisker plot describing the relationship between the mission times to reach the target and the average aggressiveness.
}
\label{Figure:Experiment.Factors.Aggresive}
\end{figure}

Furthermore,  
we explore the relationship between the aggressiveness of the vehicles and 
their performances in the simulation. Fig.~\ref{Figure:Experiment.Factors.Aggresive} shows box-and-whisker plots
that analyse the simulation results of the $6$-vehicle case. 
Fig. \ref{Subfigure:Experiment.Factors.Aggresive.Dis} illustrates the 
relationship between average aggressiveness and minimal distances. Fig. \ref{Subfigure:Experiment.Factors.Aggresive.Step} depicts the relationship between average aggressiveness and mission times. 
Generally, we observe that the evaluation of minimal distances and mission times becomes worse along with the growth of aggressiveness. 
However, the degradation caused by the growth of aggressiveness is mild. Especially, when the average aggressiveness is larger than $0.3$,
the minimal distances, as well as the missions times, do not change much. 
These results indicate that aggressive vehicles take sufficient time to decide the actions that ensure the safety requirement.

\subsubsection{Interaction between vehicles}
\label{Subsubsection:Experiments.Analysis.Details}
We examine the actions of the vehicles at each time step and 
analyse the interactions of different vehicles from the following two aspects.

\begin{enumerate}[1)]
\item The accuracy of prediction of other vehicles' acceleration.

\item The behaviours of the vehicles at each time step. 
\end{enumerate}

\begin{figure}
\centering 
{\includegraphics[height=0.2\textwidth]{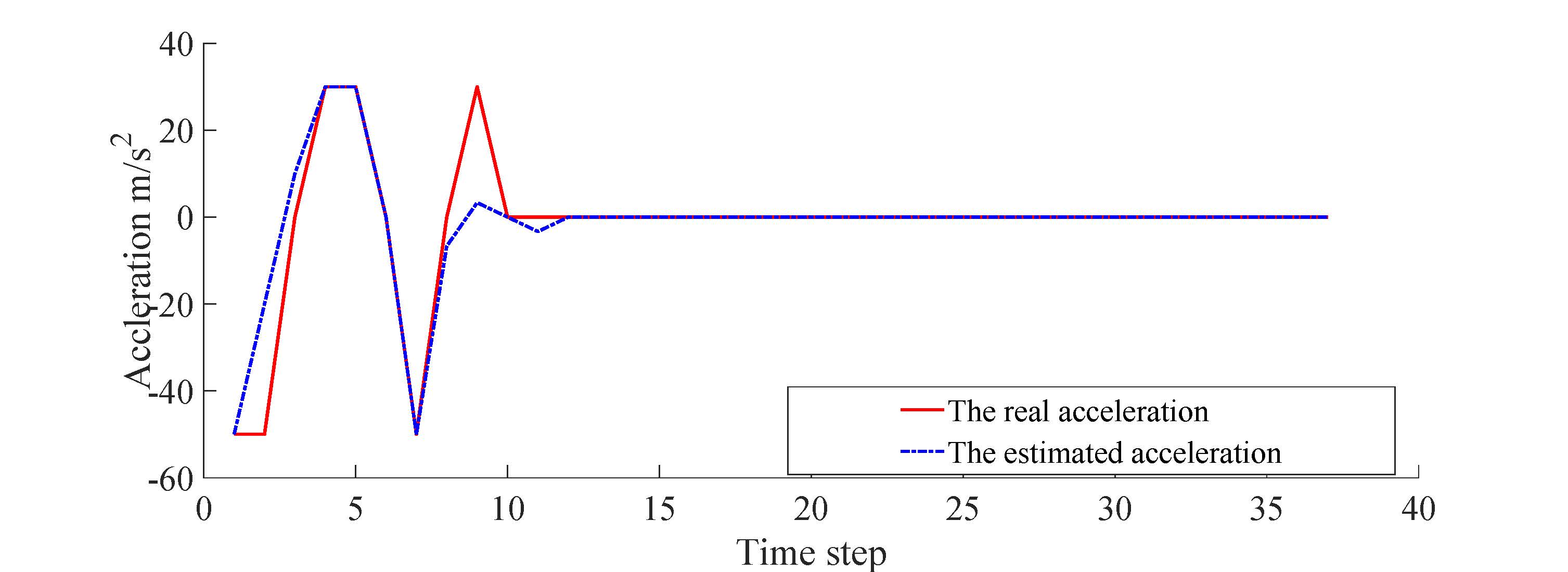}}
 \caption{The acceleration of a particular vehicle and the predicted acceleration by one of its neighbour in a representative simulation involving eight vehicles.
The $x$-axis denotes the time steps, while the $y$-axis denotes the acceleration.
 The solid {\color{red}red} line denotes the real acceleration of the vehicle, while the {\color{blue}blue} dashed line denotes the acceleration predicted by one of its neighbours.}
 \label{Figure:Experiement.Estimation}
\end{figure}

\begin{figure}
\centering    
 \subfigure[]
 {\label{fignavex}
  \includegraphics[height=0.21\textwidth]{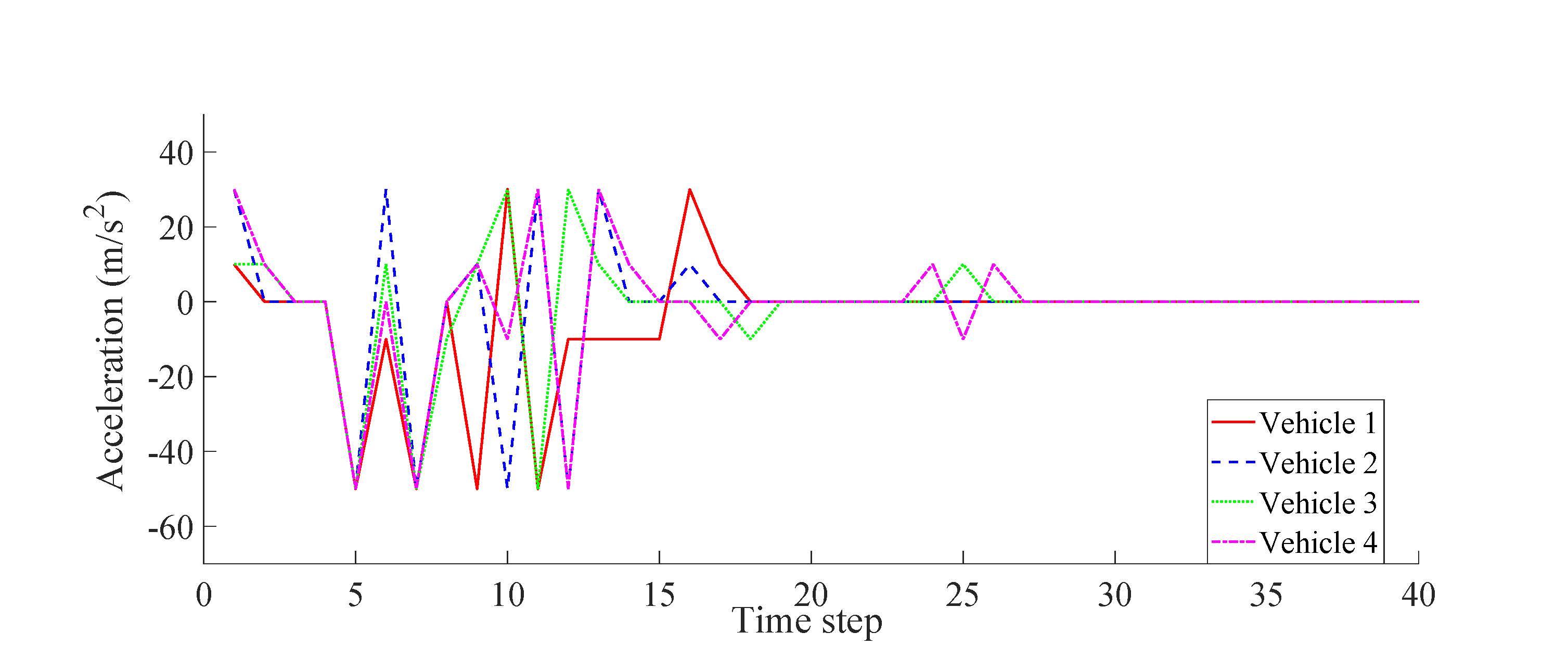} 
  }\\
  \subfigure[ $t \leq 6$]
 {\label{SubFigure:Experiment.Instance0}
  \includegraphics[width=2.5cm]{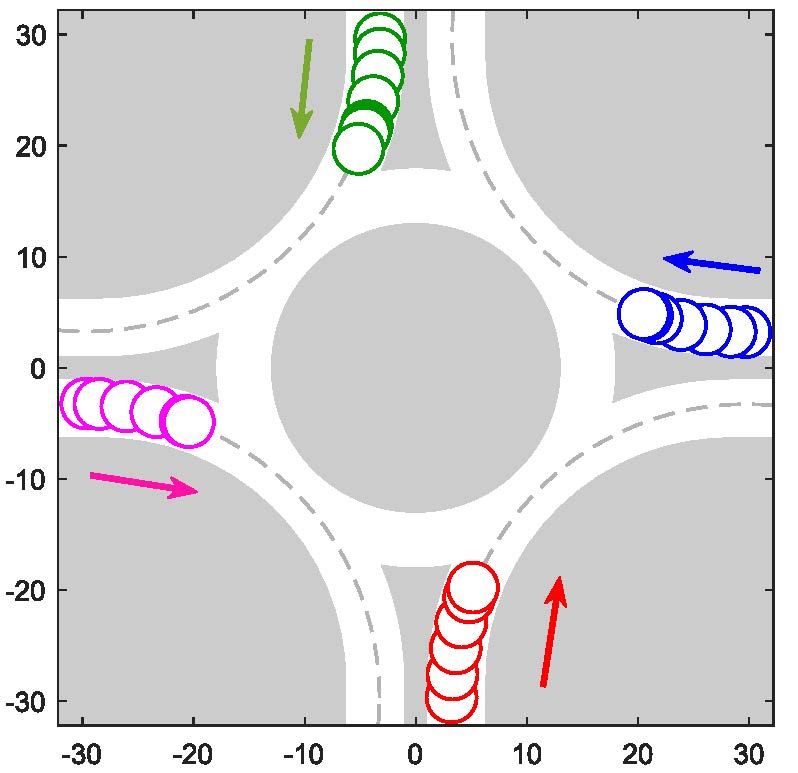}
  }
  \subfigure[$7\leq  t \leq 15$]
  {\label{SubFigure:Experiment.Instance1}
  \includegraphics[width=2.5cm]{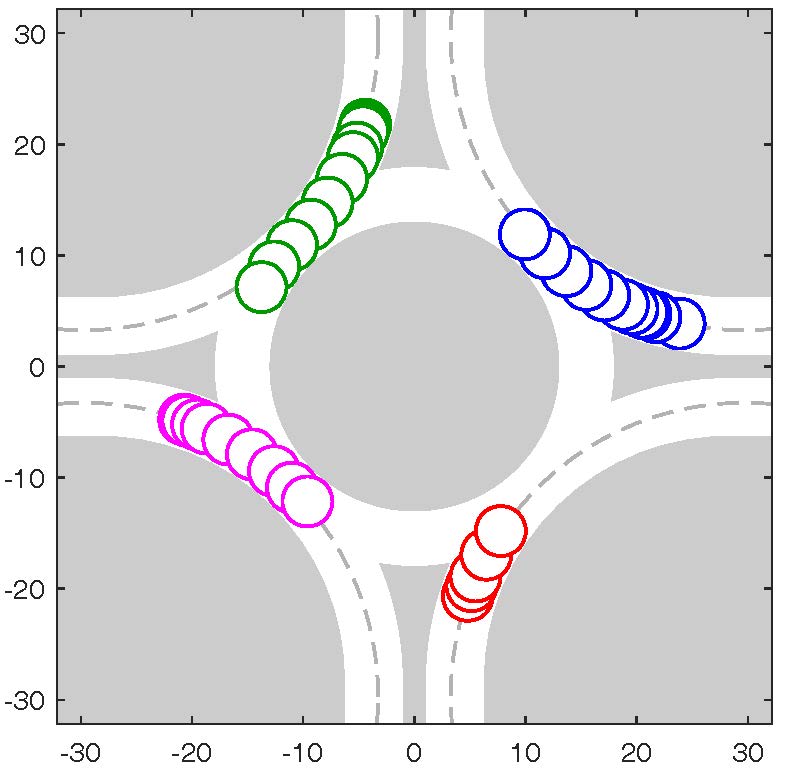}}
   \subfigure[  $t \geq16$]
  {\label{SubFigure:Experiment.Instance2}
  \includegraphics[width=2.5cm]{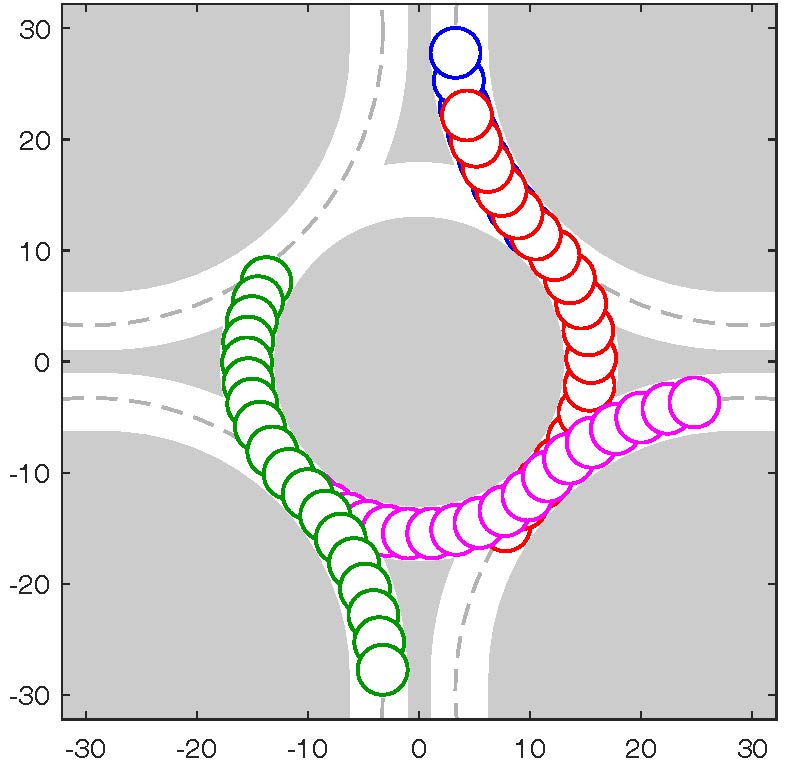}}
   \caption{(a) The interaction among four vehicles during a representative simulation.
The x-axis denotes the time steps, while the y-axis denotes the acceleration. Each line denotes the acceleration of each vehicle from the beginning to the end of the simulation.\\ 
   (b)--(d) Positions of the four vehicles at different time steps. The circles with different colours represent the position of different vehicles at each time step. (b), (c), and (d) illustrate the scenes scenes during the time-step interval in  $[0, 6]$, $[7, 15]$, and $[15, 32]$, respectively.
   The acceleration of each vehicle in (b)--(d) is shown as the line with the same colour in (a).
}
\label{Figure:Experiment.Instance}
\end{figure}


Fig. \ref{Figure:Experiement.Estimation} shows the acceleration of a particular vehicle (the solid {\color{red}red} line) and the predicted acceleration (the {\color{blue}blue} dashed line) by one of its neighbour in a representative simulation involving eight vehicles.
In the beginning,
the prediction (based on Nash equilibria in Section~\ref{section:decision}) 
 has some errors. However, 
as the neighbour observes more behaviours of the particular vehicle, the prediction becomes more accurate.
Thus, we can infer that our update process in Section~\ref{section: update} assists vehicles to
 predict their neighbours' behaviours efficiently.

Furthermore, 
in Fig.~\ref{Figure:Experiment.Instance},
we present the interactions among four vehicles during a representative simulation.     
Each line in Fig.~\ref{fignavex} illustrates the  
acceleration of each vehicle. 
Fig.~\ref{Figure:Experiment.Instance}(b)--(d) illustrate the positions of the four vehicles during three different time intervals. 
Notice that, 
the four vehicles frequently interact with each other, and each vehicle changes its decision in response to other vehicles' behaviours.  
In the beginning, the vehicles tend to decelerate when they enter the roundabout, in order to avoid the potential risks.
Then, the vehicles update their estimated aggressiveness of the other vehicles through their observations and decide appropriate strategies. 
 The experimental results show that
 the vehicles can safely interact in the intersection by making rational decisions.

\section{Conclusion}
{ 
We propose a distributed decision-making process for multiple autonomous vehicles at a roundabout.
Our process balances safety and speed of the vehicles.
Using the proposed concept of \emph{aggressiveness}, we formulate the interactions between the vehicles as finite sequential games.
The decision making, as well as the predictions of future configurations and estimations of parameters of other vehicles, is based on Nash equilibria of these sequential games.
We demonstrate the performance of our approach by performing numerical simulations, showing the feasibility and the trade-off between safety (collision rate, minimum distance) and speed (average mission time) optimisations. }

\section*{Acknowledgment}
The authors are supported by ERATO HASUO Metamathematics for Systems Design Project (No. JPMJER1603), JST. J. Dubut is also supported by Grant-in-aid No.~19K20215, JSPS.





%
%

\newpage
\section*{Appendix: 1-round sequential game with perfect information} 
\label{sec:appendix}

In this section, we give a detailed explanation of the sequential game in Section~\ref{section:decision}.
At each time step $t$,
each vehicle $i$ considers a \emph{$1$-round sequential game}
\begin{equation*}
G_i(t)=(\NEI_i(t), \Sigma, [\iest{K}{j}{t}]_{j\in\NEI_i(t)}, \prec),
\end{equation*} 
where $\NEI_i\tbrac{t}$ is the set of players, $\Sigma$ is the set of all possible strategies for each player, and $\iest{K}{j}{t}$ is the function in Eq.~\eqref{eq:gamecost}, and $\prec$ is an order of players.


Let $\NEI_i\tbrac{t} = \{ p_1, \ldots, p_{\lvert \NEI_i\tbrac{t} \rvert}\} $.
A vector 
$\begin{bmatrix}
\sigma_{p_1} & \ldots & \sigma_{p_{\lvert \NEI_i\tbrac{t} \rvert}}
\end{bmatrix}$
$ 
\in \Sigma^{\NEI_i\tbrac{t}}$ of strategies is called a \emph{best response} of player $p_k$, $k \in \{1, \ldots, \lvert \NEI_i\tbrac{t} \rvert\}$, if
\begin{align*}
    \iest{K}{p_k}{t}&(
    \begin{bmatrix} 
    \sigma_{p_1} & \ldots & \sigma_{p_k} & \ldots & \sigma_{p_{\lvert \NEI_i\tbrac{t} \rvert}}
    \end{bmatrix}
    )\\
    &\leq 
    \iest{K}{p_k}{t}(
    \begin{bmatrix} 
    \sigma_{p_1} & \ldots & \sigma_{p_k}' & \ldots & \sigma_{p_{\lvert \NEI_i\tbrac{t} \rvert}}
    \end{bmatrix}
    )
\end{align*}
for any strategy $\sigma_{p_k}' \in \Sigma$.
The strategy profile is also called a \emph{Nash equilibrium} if it is a best response of all players.
In other words, a Nash equilibrium is a strategy profile such that no player can reduce her cost by changing her strategy, provided that all other players do not change theirs.

The players take turns selecting their strategies according to the order $\prec$ -- meaning that if $p_i \prec p_j$ then $p_j$ selects her strategy before $p_i$ -- and the game stops after all players have selected their strategies.
A sequential game $G$ is a game with \emph{perfect information} if each player remembers the history of all strategies played before her.
In Section~\ref{section:decision}, this order $\prec$ is determined by the estimations of aggressiveness.

\begin{figure} [t]
\begin{center}
\includegraphics[width=0.4\textwidth]{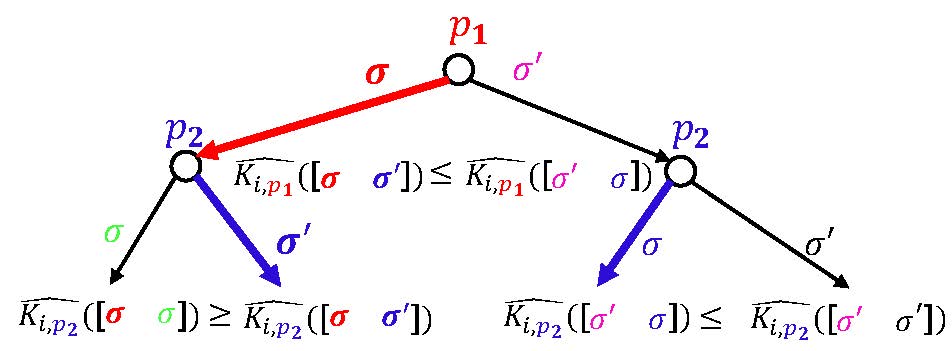}
\end{center} 
\caption{\label{fig:extensive_form}An extensive-form of a sequential game $G_i(t)$ played between $p_1$ and $p_2 \in \NEI_i(t) = \{ p_1, p_2\}$, where $p_2 \prec p_1$ and $ \Sigma = \{\sigma,\sigma'\}$.  
Notice that if $p_1$ selects strategy ${\color{red}\bm{\sigma}}$ and $p_2$ selects strategy ${\color{blue}\bm{\sigma'}}$, then $p_1$ (\emph{resp.} $p_2$) cannot reduce her cost by
 changing her strategy from ${\color{red}\bm{\sigma}}$ to ${\color{magenta}{\sigma'}}$ (\emph{resp.} from ${\color{blue}\bm{\sigma'}}$ to ${\color{green}{\sigma}}$) as long as the other player does not change her strategy.
Therefore, the vector of strategies $[ {\color{red}\bm{\sigma}} ~~~ {\color{blue}\bm{\sigma'}}]$ is a Nash equilibrium of this game.} 
 \end{figure}

The extensive-form of a $1$-round sequential game with perfect and complete information can be described as a finite decision tree.
Fig.~\ref{fig:extensive_form} shows an extensive-form of such a game played between two players.
As we assume that $p_2 \prec p_1$, $p_1$ selects her strategy before $p_2$ at the root of the tree.
We can obtain a Nash equilibrium of the game by applying the backward induction algorithm on the decision tree.
For the game in Fig. \ref{fig:extensive_form}, we can compute a Nash equilibrium as follows.
First, we compute best responses for $p_2$ in both subtrees, which represent the cases that $p_1$ selects ${\color{red}\bm{\sigma}}$ and ${\color{magenta}{\sigma'}}$.
If $p_1$ selects ${\color{red}\bm{\sigma}}$ (\emph{resp.} ${\color{magenta}{\sigma'}}$), then $\begin{bmatrix}
{\color{red}\bm{\sigma}} & {\color{blue}\bm{\sigma'}}
\end{bmatrix} $ 
(\emph{resp.} 
$\begin{bmatrix}
{\color{magenta}{\sigma'}} & {\color{blue}{\sigma}}
\end{bmatrix}$)
is a best response for $p_2$.
We then compute the best response for $p_1$ by considering the best responses of $p_2$ for each subtree. In this case, $\begin{bmatrix}
{\color{red}\bm{\sigma}} & {\color{blue}\bm{\sigma'}}
\end{bmatrix} $ is a Nash equilibrium. 
 
The extensive-form of the game is a tree with the branching of $\lvert\Sigma\rvert$ and is $\lvert\NEI_i\tbrac{t}\rvert$ deep. Consequently, it has $O(\lvert\sigma\rvert^{\NEI_i\tbrac{t}})$ nodes. Each node requires the computation of an accumulated cost. The complexity of this computation is $O(h\cdot\lvert\NEI_i\tbrac{t}\rvert)$, because, for every time step until the horizon, we have to compute $i^+$ and $i^-$. In conclusion, the complexity of this backward induction is $O(h\cdot\lvert\NEI_i\tbrac{t}\rvert\cdot\lvert\Sigma\rvert^{\NEI_i\tbrac{t}})$.

 We invite interested readers to see 
textbooks (e.g.~\cite{O2009}) for more details on game theory.

\end{document}